# Work-Function-Resolved Imaging of Relaxation Oscillations and Chemical Spillover in CO Oxidation over Platinum Surfaces

Karel Vařeka[1], Michal Potoček[1,2], Adam Očkovič[2], Tomáš Šikola[1,2], Zhu-Jun Wang[3], Petr Bábor[1,2], Miroslav Kolíbal[1,2*]

[1]Brno University of Technology, Central European Institute of Technology, Purkyňova 123, 61200 Brno, Czech Republic

[2]Brno University of Technology, Faculty of Mechanical Engineering, Institute of Physical Engineering, Technická 2, 616 69 Brno, Czech Republic

[3]School of Physical Science and Technology, Shanghai Tech University, Shanghai, 201210, China

[*]kolibal.m@fme.vutbr.cz

ABSTRACT

Chemical waves of CO oxidation on platinum surfaces exhibit complex spatio-temporal self-oscillations, yet the local electronic mechanisms driving their propagation remain poorly understood under operando conditions. In this work, we combine operando scanning electron microscopy with frequency-modulated Kelvin probe force microscopy (FM-KPFM) to simultaneously map secondary electron contrast and local work-function variations during CO oxidation on Pt. By utilizing the KPFM tip as a localized sensor, we provide the first work-function-resolved imaging of reaction fronts, enabling an unambiguous physical assignment of CO- and oxygen-covered states. Our results demonstrate that the spillover process of chemical wave—the transition and expansion of adsorbate phases—is characterized by a pronounced temporal asymmetry and spatial heterogeneity transition thresholds. KPFM identifies a rapid onset of oxygen coverage followed by a gradual, diffuse relaxation back to the CO-covered state, indicative of relaxation-type oscillations even at low pressures ($10^{-2}$ Pa). Correlative reaction-diffusion simulations reproduce this wave morphology, confirming that the high-resolution work-function signal provides unique insights into the internal structure and kinetic heterogeneity of the working catalyst surface.

*Keywords:*

Chemical wave, Spillover, CO oxidation, Scanning electron microscopy, Kelvin probe force microscopy, Relaxation oscillations, Catalytic self-organization

INTRODUCTION

Heterogeneous catalysis underpins many of the chemical transformations essential for modern society, including large-scale industrial synthesis, environmental remediation, and energy technologies. Catalysts are inherently complex systems whose activity, selectivity, and stability depend on the intricate interplay between their surface structure, electronic properties, and adsorbate coverage. It has been well understood that such knowledge becomes available only through *in situ* studies of these reactions. Heterogeneous gas-phase catalysis has been studied by ultra-high vacuum-based surface science techniques (e.g., LEED, XPS, PEEM, STM, etc.) with great success.[1–8] Although these studies had a tremendous impact on our mechanistic understanding of the reactions on the catalysts' surfaces, the existence of a pressure gap between these studies and technologically relevant conditions[9,10] have initiated an extensive research and development of operando modification of these techniques. Currently, most of the techniques that provide microscopic (TEM, STM) and spectroscopic (XPS) data can be performed close to *operando* conditions.[1,11–16] Yet, just recently,[17] a possible discrepancy between microscopic and spectroscopic data has been pointed out: a focus on a selected small feature as in the former and averaging the signal from large areas on the sample in the latter, does not allow for distinguishing between active and inactive catalyst sites. While looking at one particular site, doesn't the reaction happen at the neighboring one, out-of-sight? Do we observe the catalyst in action or rather its reaction to varying environments due to reactions happening somewhere else? In response, a great effort has been raised recently to develop spectroscopies with high lateral resolution,[18,19] new signal detection techniques[15,20,21] and, finally, correlative methodologies, where the data from various techniques can be compiled into a site-specific multidimensional view of the working catalyst.[19,22]

Here, we support this endeavor by combining two techniques that have been integrated to track reaction dynamics in situ: scanning electron microscopy (SEM) and Kelvin probe microscopy (KPFM). Secondary electron images in SEM provide a complex view of the chemical state of the surface and topography. Compared to LEEM/PEEM, the technique is not limited in terms of sample size and shape and allows imaging up to very high working pressures.[23] KPFM is a variant of atomic force microscopy (AFM) that enables local mapping of surface potential and work function differences with a nanometer-scale spatial resolution by measuring the contact potential difference between a conductive AFM tip and the sample surface. While macroscopic Kelvin probe measurements yield spatially averaged information,[24,25] the KPFM tip acts as a localized sensor capable of capturing the discrete behavior of chemical wave spillover as it propagates across the catalyst surface. The method has been successfully applied to several catalytic and catalytically relevant systems, including metal and oxide surfaces as well as supported metal nanoparticles, predominantly under inert or quasi-static conditions.[26–28] To the best of our knowledge, KPFM has not yet been employed to probe local work function changes under operando conditions in dynamically evolving catalytic systems. The integration of SEM with KPFM combines a wide-area imaging of dynamic surface contrast with a localized surface potential measurement. We demonstrate that simultaneous KPFM measurements provide provide a unique work-function signal that allows us to probe the mechanism of the chemical wave spillover process with even higher spatial and chemical sensitivity than SEM.

We apply this instrumental toolbox, together with spatially-resolved mechanistic modelling, to carbon monoxide (CO) oxidation to $CO_2$ on platinum surfaces, a system exhibiting remarkably rich kinetic behavior. On one hand, its relevance lies in its role in automotive exhaust purification and methanol synthesis.[29] On the other hand, its scientific appeal is hidden in its conceptual simplicity - involving only small molecules and a few intermediates - yet exhibiting remarkably rich kinetic behavior.[30] Early ultra-high vacuum (UHV) studies on single-crystal Pt surfaces revealed that CO and $O_2$ competitively adsorb, forming dynamic surface phases such as chemisorbed carbon monoxide and oxygen,[2] or potentially platinum oxides,[31–33] and subsurface oxygen.[33] The reaction is known to produce propagating reaction fronts and periodic oscillations that rely on the feedback between surface coverage and reactivity, as shown in the pioneering studies of G. Ertl and his co-workers.[34,35] These oscillations arise from the coupling of surface adsorption–desorption kinetics with structural transformations and possibly oxide and subsurface oxygen formation (though the community is not in agreement with the latter).

Traditionally, these low-pressure oscillations have been described by harmonic functions in PEEM studies.[2,30] However, by combining in situ SEM with localized FM-KPFM to track the reaction fronts, we have captured the complex work-function-resolved signature of the chemical wave spillover process. Our integrated measurements revealed the presence of relaxation-type oscillations even at $10^{-2}$ Pa, a regime which was previously identified only at higher mbar pressures. This unique KPFM signal provides a direct electronic fingerprint of the surface, allowing us to probe the internal structure and threshold behavior of the spillover process with a sensitivity that surpasses traditional intensity-based imaging. Understanding the dynamics of the rate oscillations under steady external conditions provides information on the role of metastable structures, which are often critical for catalytic function.[36] Yet, despite decades of research, the precise nature of these surface patterns - their composition, structure, and the origin of contrast - remains incompletely understood. By employing *in situ* SEM monitoring with simultaneous KPFM, our study contributes to a more detailed understanding of the mechanisms underlying chemical wave spillover and pattern evolution.

RESULTS

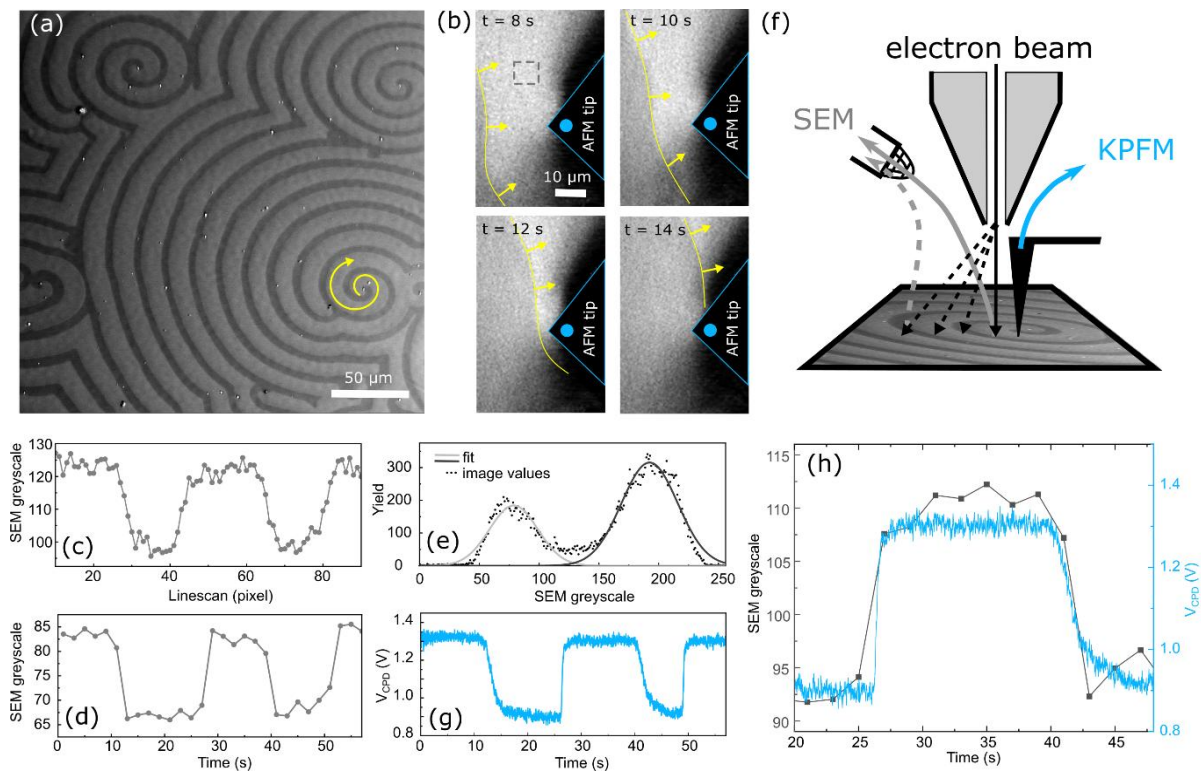

Fig. 1: Correlative SEM and FM-KPFM imaging of spatio-temporal reaction patterns during CO oxidation on Pt surfaces. (a) *In situ* SEM image showing spiral and planar reaction waves on Pt(110) during catalytic CO oxidation in Tescan UHV-SEM. Experimental conditions: $p(CO) = 4\times10^{-4}$ Pa, $p(O_2) = 1.5\times10^{-3}$ Pa, T = 170 °C . The patterns consist of two distinct greyscale levels corresponding to different surface chemical states (see Movie M1). (b) Sequence of SE images showing the evolution of a reaction front within a single grain on a polycrystalline Pt sample in the vicinity of the AFM tip (blue, see further) in FEI Versa 3D SEM. Experimental conditions: $p(CO) = 3\times10^{-3}$ Pa and $p(O_2) = 1.6\times10^{-2}$ Pa, T = 216 °C. The yellow line marks the position and the direction of the reaction front; yellow arrows show the propagation direction. See full Movie M2. (c) Spatial line scan across propagating reaction waves in Tescan UHV-SEM, total distance is 41 µm. (d) Time-resolved SEM signal acquired with a stationary electron beam (position of the beam marked by a dashed square in (b)). (e) Histogram of SEM greyscale values from a representative snapshot in (b), fitted with two Gaussian components, confirming the bimodal intensity distribution. (f) Schematic of the experimental configuration combining SEM and frequency-modulated Kelvin probe force microscopy (FM-KPFM). (g) Time trace of the contact potential difference ($V_{CPD}$) measured with a stationary AFM tip during the ongoing reaction in (b). (h) Simultaneously recorded SEM greyscale (grey squares) and $V_{CPD}$ (blue curve) signals, demonstrating direct correlation between secondary electron contrast and surface potential variations associated with CO- and O-covered regions.

The SEM image in Fig. 1a (accompanied by Supporting Movie M1) illustrates the emergence and evolution of spatio-temporal patterns on a Pt(110) single crystal during the catalytic oxidation of CO. A characteristic elliptical spiral pattern, pinned to a surface defect, coexists with propagating planar waves of two distinct brightness levels. The patterns are stable on the micrometer length scale (tens of micrometers) and evolve on a time scale of minutes.

On polycrystalline Pt samples, the spatio-temporal patterns exhibit grain-dependent morphologies and kinetics due to variations in catalytic activity and activation barriers associated with different crystallographic orientations of each individual grain [34,35]. The characteristic morphologies and kinetics of the reaction fronts remained stable under continuous imaging and showed no deviation in front propagation or pattern geometry immediately after unblanking the electron beam. An exemplary image sequence in Fig. 1b, acquired on a single grain within a polycrystalline Pt, shows a dark-contrast wave with a curved wavefront, propagating from left to right. In an attempt to probe the dynamics within individual surface waves, we employed a stationary electron beam and recorded the detector signal while the reaction proceeded. Both spatial line scans across the patterns (Fig. 1c) and time-resolved point measurements at a fixed beam position (Fig. 1d) show signal variations between two well-defined greyscale levels, separated by relatively narrow transition regions. The histogram of a representative SEM image (Fig. 1e), fitted by two Gaussian components, further confirms the two-level character of the detected signal in SEM.

Work-function-sensitive PEEM studies[2,34] have established that oxygen-covered regions appear dark due to the oxygen-induced surface dipoles and the associated increase in WF, whereas CO-covered Pt regions, characterized by a lower WF, appear bright. In SEM, however, the interpretation of SE image is more complex. The secondary electron (SE) yield arises from a complex interplay of SE generation, transport to the surface, and escape probability. For a planar metallic surface covered by adsorbates or a thin surface oxide, the dominant contribution can be attributed to local variations in the surface work function (WF), which directly affect the SE escape probability.[42,43] In addition to the influence of work function and intrinsic effects of SE generation, the detected SE signal is significantly affected by factors such as detector type and position, local electric fields near the sample, and any applied sample bias. Under certain conditions, the SE contrast can even reverse; for example, if the local electric field near the active reaction site is distorted by the presence of the conducting AFM tip (see Supporting Information, Movie M3). As a result, unambiguously matching a surface state to the observed contrast, as is possible in PEEM, becomes challenging.

To address these limitations and uncertainties in contrast identification, we implemented operando platform combining frequency-modulated KPFM (FM-KPFM) within the SEM (Fig. 1f), enabling in situ tracking and direct correlation between SE contrast and local surface potential variations. With a stationary AFM tip acting as a localized sensor, we measured the contact potential difference ($V_{CPD}$) via a bias feedback loop while simultaneously performing SEM imaging on the same surface region. This simultaneous acquisition provides a quantitative work-function signal that serves as a direct electronic signature of the surface states. Figure 1b illustrates the spatial evolution of a reaction front near the AFM tip, confirming that both SEM and FM-KPFM probe the same local area. The $V_{CPD}$ time trace (Fig. 1g) shows two distinct levels separated by $0.36 \pm 0.04$ V. Since $V_{CPD}$ directly reflects surface work function variations, the higher $V_{CPD}$ level corresponds to oxygen-covered regions and the lower to CO-covered regions, consistent with PEEM results that associate higher work function (and thus lower photoelectron emission) with oxygen-covered areas and lower work function with CO-covered areas.[30,44] The simultaneous acquisition of SEM greyscale and $V_{CPD}$ signals (Fig. 1h) thus provides an unambiguous correlation between SE contrast and the local chemical state of the surface.

Furthermore, the FM-KPFM measurements reveal a pronounced asymmetry in the reaction fronts that is not apparent in the SE signal alone, indicating that SE imaging alone may not be sufficient for identifying surface chemical states. FM-KPFM time trace shows that the transition from the oxygen-covered to the CO-covered state is characterized by an irregular and spatially extended boundary, while the reverse transition (CO to O) forms a sharp and well-defined interface. This intrinsic asymmetry of the reaction fronts is directly accessible only through the quantitative surface potential measurements provided by FM-KPFM.

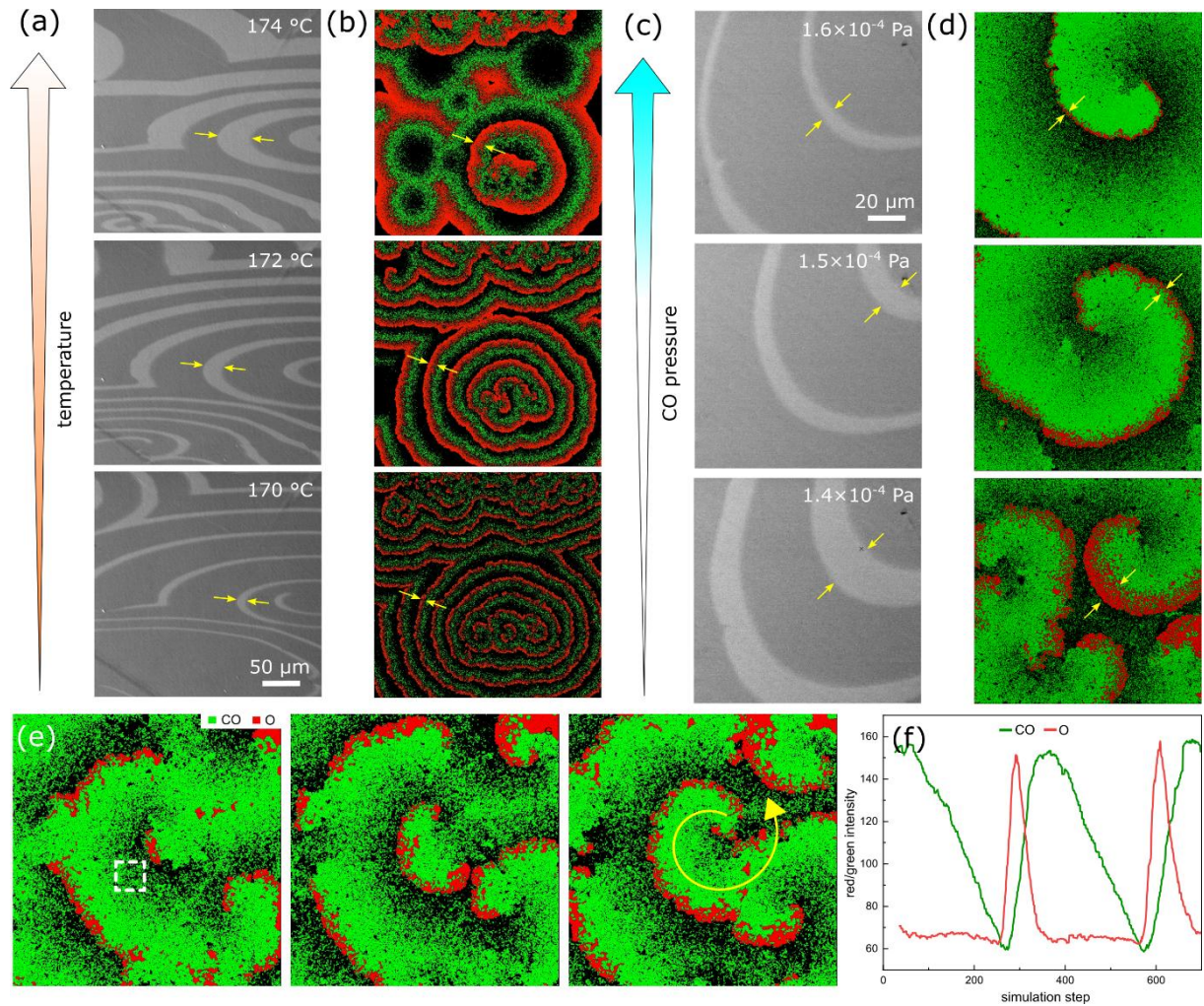

Fig. 2: Reaction-diffusion model outcomes compared with SEM experiments. (a) SEM and (b) simulation snapshots of CO (SEM: dark, simulation: green) and oxygen (SEM: bright, simulation: red) waves on platinum (simulation: black) taken for different sample temperatures. Experimental conditions: $p(CO) = 7.0 \times 10^{-4}$ Pa and $p(O_2) = 4.1 \times 10^{-3}$ Pa. (c) SEM and (d) simulation snapshots taken for different CO partial pressures. Experimental conditions: $p(O_2) = 3.4 \times 10^{-3}$ Pa, T ~ 180 °C. (e) Snapshots from a continuous movie (see Supporting Information, Movie M4) showing the spiraling pattern formation. (f) Time-resolved simulation signal (x axis showing simulation steps, i.e., time), with separate red (O) and green (CO) intensities, as read-out from a dashed square in (e)).

To corroborate the experimental observations, we have employed a spatially resolved reaction–diffusion Metropolis Monte Carlo (MMC) simulation model.[37] The simulations reproduce the experimentally observed spatio-temporal pattern formation, e.g. its periodicity. The periodicity and relative width of the waves are controlled by an interplay between adsorption–desorption kinetics and the nonlinear reaction terms. The numerical results qualitatively capture the dependence of pattern morphology on external parameters such as temperature and the $CO/O_2$ partial pressure ratio. Fig. 2a shows SEM micrographs acquired at increasing sample temperature, revealing a progressive widening of oxygen-covered (bright) waves at the expense of CO-covered (dark) waves. This trend reflects the shift in competitive adsorption towards preferring oxygen at higher sample temperature, consistent with previous reports.[35,45] The corresponding simulations in Fig. 2b reproduce this behavior: increasing sample temperature, implemented via modified rate constants following Arrhenius equation, leads to a broader oxygen wave segments and reduced CO wave width.

A similar qualitative agreement is obtained when varying the $CO/O_2$ pressure ratio. In the experiment (Fig. 2c), increasing CO partial pressure leads to a widening of CO-covered regions. The simulations (Fig. 2d) reproduce this trend, confirming that the model captures the essential dependence of the pattern morphology on external parameters. While a quantitative comparison would require explicit conversion between model reaction rates (in Langmuir units) and experimental partial pressures (Pa) -- which is not the scope of this work -- the qualitative agreement demonstrates that the model reliably reproduces the key dynamical features of the system, including the formation and rotation of spiral waves (Fig. 2e). The spiral core, highlighted in Fig. 2e, acts as a persistent source of propagating reaction fronts, similar to the experimental observation (ref.[25] and Fig. 1a).

Importantly, the simulations provide insight into the origin of the temporal asymmetry observed experimentally in the KPFM measurements. Extracting time-resolved point signals from the model (Fig. 2f), analogous to the SEM and KPFM experiments, reveals a rapid onset of oxygen coverage followed by a transition from O-to CO-coverage. This asymmetry, characterized by a sharp reaction front and a slow recovery tail, reveals the internal structure of the chemical wave spillover event. The transient region then transforms into a gradual decay of the CO-related signal. This intrinsic asymmetry of the reaction front dynamics is clearly consistent with the KPFM signal, which is directly sensitive to work function changes associated with surface coverage. In contrast, the SEM signal does not resolve this asymmetry, as the secondary electron contrast lacks sufficient sensitivity to the transient coexistence regime. Improving the signal-to-noise ratio of SE imaging is challenging, as longer dwell times hinder tracking of fast-moving fronts, while increased beam currents enhance the risk of beam-induced artifacts such as hydrocarbon dissociation and catalytic poisoning. KPFM signal thus provides a greater detail of wave spillover dynamics than SEM.

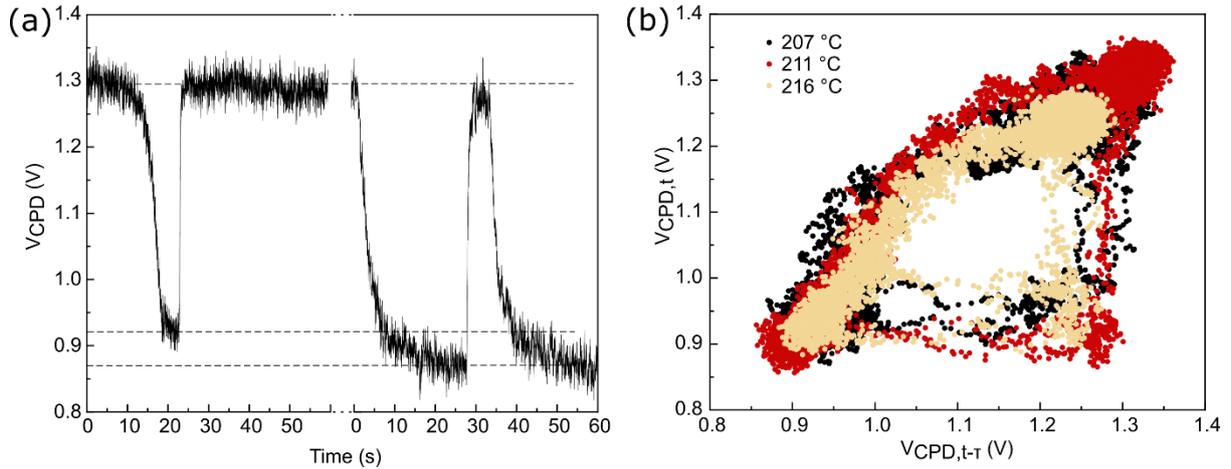

Fig.3: Various KPFM data analysis. (a) Time-resolved analysis, showing that the potential differences vary depending on the sample position. Experimental conditions: $p(CO) = 3\times10^{-3}$ Pa, $p(O_2) = 1.4\times10^{-2}$ Pa, T = 214 °C. (b) phase-portrait of KPFM data at different temperatures (T = 207 °C, 211 °C and 216 °C.), where $\tau = 0.9$ s is the time delay (equal for all datasets). Experimental conditions: $p(CO) = 3\times10^{-3}$ Pa, $p(O_2) = 1.5\times10^{-2}$ Pa.

The observed similarity in asymmetrical decay between the simulated time traces (Fig. 2f) and the FM-KPFM measurements (Fig. 1g) prompted a more detailed analysis of the experimental KPFM data. We found that the low work-function (WF) state in the FM-KPFM time traces exhibits distinct absolute values even under otherwise constant external conditions (Fig. 3a). In our experiments, the maximum potential difference between the high- and low-WF states has reached 0.46 eV. Notably, the transition to the high-WF state does not occur at a unique or characteristic value of the low-WF potential, indicating the absence of a single well-defined threshold in $V_{CPD}$ for switching the chemical state of the surface to O-covered state.

To further analyze the transition dynamics, we constructed phase portraits from the KPFM time traces using the time-delay embedding method, as introduced previously in.[2] In this representation, the signal $V_{CPD}(t)$ is plotted against $V_{CPD}(t-\tau)$, where the same delay time $\tau$ was used for all datasets. The resulting phase portrait (Fig. 3b), containing data recorded at three different temperatures, exhibits a pronounced triangular shape. Most data points cluster near three corners and along a diagonal connecting two of them, in clear contrast to the elliptical phase portraits previously reported from PEEM measurements.[2] The triangular geometry reflects the asymmetry of the oscillations observed in time traces (Fig. 3a), with a slow evolution along the hypotenuse and rapid switching between the metastable states along the steep sides. The asymmetry becomes more pronounced with increasing temperature, as the phase portrait progressively approaches a right-angled triangular shape and the data points cluster more closely along the linear branch connecting the corners. We emphasize that even subtle temperature variations (<10 °C) lead to clearly distinguishable changes in the phase portrait, highlighting the high sensitivity of the oscillatory dynamics to external conditions.

DISCUSSION

The combined SEM–FM-KPFM approach enables an unambiguous physical assignment of the observed contrast to the underlying surface chemistry. While SEM alone reveals two distinct intensity levels associated with propagating reaction fronts (Fig. 1c-e), only the simultaneous KPFM measurement directly links these states to local work-function variations. The higher work-function state is consistently associated with oxygen-covered Pt, whereas the lower work-function state corresponds to CO-covered regions, in agreement with earlier PEEM studies of CO oxidation on Pt surfaces.[2] Importantly, the maximum measured contact potential difference (up to 0.46 eV) rather argues against formation of a condensed, large scale transient surface oxide or subsurface oxide phases under the present reaction conditions. A transition from chemisorbed oxygen to oxide involves substantial structural and electronic rearrangement typically associated with larger surface dipole moments and, thus, larger work function changes than observed here.[46] In contrast, the measured $V_{CPD}$ values lie within the range previously reported for CO oxidation on Pt at even lower pressures, where oxide formation is considered unlikely.[47] While we cannot fully exclude presence of isolated clusters of transient oxides[36] that may remain below the spatial resolution of FM-KPFM, our data suggest that an extended oxide phase is unlikely to govern the observed front dynamics. Instead, the oxygen species participating in the reaction correspond predominantly to chemisorbed O on metallic Pt, acting as the main reactive partner in $CO_2$ formation. The reaction is consistent with proceeding via the classical Langmuir--Hinshelwood mechanism rather than an oxide-mediated (Mars-van Krevelen) pathway.

However, within a Langmuir–Hinshelwood scenario, one might in principle expect three work-function levels, corresponding to O-covered Pt, CO-covered Pt, and clean reconstructed Pt. Based on the profile of our KPFM transition curves, we hypothesize that the slow transition between O-covered and CO-covered phases in our profile converges toward the clean Pt state, which forms the baseline of $V_{CPD}$ measurements. However, the close proximity of the work functions of clean and CO-covered Pt (below 0.2 eV for Pt(100), ref[47]) prevents distinguishing between the two states. In addition, as discussed below, the KPFM data reveal a gradual decrease of CO coverage along the propagating CO wave, resulting in a smooth work-function gradient that further blurs any sharp distinction between clean and CO-covered Pt.

Both KPFM time traces and simulations reveal a pronounced asymmetry of the reaction fronts, characterized by a rapid onset of the oxygen-covered state and a slower recovery of the CO-covered state. This steep leading edge of the oxygen wave reflects the strongly nonlinear kinetics of dissociative $O_2$ adsorption once CO coverage locally drops below a critical level.[48,49] However, Fig. 3a clearly demonstrates that this critical level is not a single global threshold: the transition to the high work-function (oxygen-covered) state occurs at different absolute values of the low work-function state, even under steady external conditions. This observation suggests that the effective threshold for oxygen adsorption depends on the local kinetic environment, including coverage fluctuations, surface reconstruction state, and possibly defect density. While mean-field descriptions of CO oxidation often invoke a well-defined global critical CO coverage,[2] spatially resolved experiments have previously shown that reaction fronts nucleate locally and propagate depending on local conditions.[34,50] Our KPFM data provide evidence of such a locality at the level of work-function-resolved coverage dynamics.

Both the reaction–diffusion simulations and the KPFM measurements indicate that the propagating reaction wave is not composed of two spatially homogeneous states separated by an infinitely sharp interface. Instead, a continuous gradient of adsorbate coverage exists within each wave region, as the competition between site preference and lateral repulsion prevents the formation of perfectly ordered overlayers.[51] This is directly visible in the simulated coverage maps (Fig. 2) and reflected in the KPFM time traces (Fig. 3a), which exhibit an extended transition region rather than an abrupt switching between isolated $V_{CPD}$ levels. This finding provides additional insight compared to earlier KPFM experiments performed using macroscopic electrodes or large-area probes,[47,52] where the measured signal represents a spatial average over many domains, preventing unambiguous explanation of the $V_{CPD}$ variations. The possible existence of substructures within the adsorbate waves was investigated by Ertl and co-workers using operando diffraction[47] and later by scanning tunnelling microscopy at very low temperatures, which revealed nanoscale heterogeneities and atomic oxygen clustering.[43] By increasing the spatial resolution of Kelvin probe by using a sharp tip, and by correlation with simulations, our work reveals that the work-function profile across the wave contains internal structure related to gradient in surface coverage. In a perspective, recent developments in KPFM promise to achieve close-to-atomic resolution in the KPFM mode,[53] potentially allowing quantification of clustering within the reaction waves under operando conditions.

Given the superior lateral resolution of a typical SEM measurement, it might appear surprising that the internal structure within the surface waves is not detectable in SEM images (Fig. 1). In order to determine the theoretical capability of SEM to distinguish inner surface features, we simulated SE images of regions with different work functions using Monte Carlo–based simulations within the Nebula package,[54] while progressively reducing the size of these regions to establish the detection limit (see Supporting Information, Fig. S1). Utilizing an electron probe with 1 nm spot size, a distinct contrast between the 55 nm-wide O-covered Pt stripe and surrounding CO-covered Pt is clearly resolved in the simulated SE image with 50 µm field of view (commonly utilized to image spatio-temporal patterns) (Fig. S1c). Such a small probe size is, however, insufficient to achieve reasonable WF contrast in a real SE image. For a larger spot size, close to the real experimental conditions (40 nm), the contrast becomes significantly obscured. On a 55 nm O adsorbate island, the WF-induced contrast is only weakly detectable at 0.5 µm field of view (Fig. S1f). Furthermore, simulating alternating adsorbate stripes with lateral sizes down to 5 nm, the resulting image contrast (and related histogram) deteriorated even further (see Fig. S1m). Besides the issues with a low signal-to-noise ratio, topographic artifacts upon gas adsorption [29] induced by reaction-driven nanofaceting of Pt surface[44] may hinder the WF contrast even if the imaging conditions are further improved. This implies that while SEM offers a superior spatial resolution compared to traditional PEEM or LEED studies,[31,34,35] nm-sized features within the reaction waves remain challenging to observe in SE images.

The phase portrait derived from the KPFM time series (Fig. 3b) exhibits a distinctly triangular shape, in contrast to the elliptical phase portraits reported in earlier PEEM studies [2]. An elliptical portrait typically indicates nearly harmonic oscillations with comparable time scales for forward and backward transitions. In contrast, the triangular geometry observed here reflects relaxation-type oscillations with strongly separated time scales. The nearly linear hypotenuse is consistent with a slow evolution of the system along one metastable branch (low-reactive state), while the steep sides are consistent with rapid switching events between the CO- and O-covered states (high-reactive state). The clustering of data points in the three corners indicates prolonged residence in quasi-stationary states separated by fast transitions, consistent with excitable reaction–diffusion dynamics. Relaxation-type oscillations have been only reported in the midrange (0.1 mbar)[35] or atmospheric pressure regime,[55] but not in low-pressure conditions (~$10^{-2}$ Pa). Our data suggest that the relaxation type oscillations may be more frequent than previously considered in this type of reaction, due to a limited lateral resolution of the previously utilized techniques. Such a conclusion is further supported by a phase portrait derived from SEM data (Supporting Information, Fig. S2), which is nearly elliptical, owing to the lower resolution of the SE images.

CONCLUSIONS

In summary, the combined SEM–KPFM measurements provide a direct correlation between electron-intensity contrast and local work-function variations, enabling an *in situ* tracking of reaction fronts with unprecedented electronic detail. This instrumental integration allows us to bridge the gap between wide-area imaging and localized surface potential measurements, leading to an unambiguous assignment of the propagating states to CO- and O-covered Pt. By employing the KPFM tip as a stationary probe, we have successfully resolved the behavior and mechanism of the chemical wave spillover process as it traverses the catalyst surface. The data indicate that the CO oxidation on Pt surface under these conditions is consistent with a relaxation-type regime characterized by a pronounced time-scale separation. Specifically, the KPFM signal reveals that the spillover of the oxygen wave is not a symmetric transition but involves a sharp non-linear onset and a gradual coverage relaxation. This internal structure of the reaction front is not accessible to traditional intensity-based imaging methods like SEM alone. Furthermore, the KPFM measurements reveal that the onset of the spillover event is governed by local kinetics rather than a single global critical value. The observed deviation from an elliptical to a triangular phase portrait reflects the intrinsic relaxation character of the oscillations under our experimental conditions and highlights the enhanced sensitivity of KPFM to the internal structure of the reaction front. The measured potential differences suggest the absence of stable surface or subsurface oxide phases, supporting the interpretation that the reaction proceeds within the Langmuir–Hinshelwood framework. Overall, these findings refine the classical mean-field description of oscillatory CO oxidation by emphasizing the importance of spatially resolved kinetics and local work function-resolved spillover behavior.

MATERIALS AND METHODS

We have utilized two types of platinum substrates to host the reaction: a Pt(110) single crystal (99.999% purity, orientation accuracy < 0.1°, SPL) and polycrystalline Pt wire (100 μm diameter, 99.99% purity, GoodFellow). The polycrystalline wire was mechanically flattened to increase the available surface area for imaging. To achieve a contamination-free surface, samples underwent rigorous cleaning within an ultra-high vacuum chamber with base pressure in the range of $10^{-7}$ Pa. The cleaning procedure consisted of high-temperature annealing (1000 °C) in an oxygen atmosphere (p($O_2$) = 5×$10^{-4}$ Pa, 99.999%, Messer) to remove residual hydrocarbon species. In case of severe surface contamination, 2 kV $Ar^+$ ion sputtering followed. Finally, the samples were flash annealed up to 1200 °C. For KPFM measurements, the chamber and sample were additionally treated with oxygen plasma for 15 minutes to mitigate any ambient contamination introduced in the case of sample transfer to a different SEM setup housing KPFM.

The catalytic CO oxidation was investigated using two complementary systems: a custom-built ultrahigh vacuum scanning electron microscope (UHV-SEM) developed in collaboration with Tescan (Brno, Czech Republic), and a commercial FEI Versa 3D DualBeam (SEM/FIB) housing a NenoVision LiteScope atomic force microscope. SEM imaging was performed at an acceleration voltage of 5 kV and a beam current of 4 nA. To maintain the necessary temporal resolution for observing dynamic wavefronts, no signal averaging was applied, and a dwell time of 1-5 µs was utilized. Secondary electron (SE) signals were collected using an Ion Conversion and Electron (ICE) detector on the Versa 3D system and chamber SE detectors on UHV-SEM system. To ensure the integrity of the surface dynamics under continuous electron beam illumination, the possible electron beam influence was monitored by performing intermittent beam blanking. These experiments confirmed that the reaction fronts continued to propagate at a constant velocity regardless of presence or absence of the electron irradiation.

The formation of spatio-temporal patterns was initiated by introducing constant flow of CO (99.995%, Linde) and $O_2$ (99.999%, Messer) at a predefined ratio. The partial pressures were maintained between $10^{-4}$ Pa (CO) and $10^{-3}$ Pa ($O_2$) in case of the UHV SEM observations, and $10^{-3}$ Pa (CO) and $10^{-2}$ Pa ($O_2$) in case of the Versa SEM. Individual reaction conditions are always indicated in figure captions. The sample temperature was controlled by an in-house resistive heating holder, and monitored via an infrared pyrometer (Optris CTvideo 3MH1, emissivity set = 0.1). The absolute temperature measurement uncertainty, estimated as ±20 °C, arises from an inherent uncertainty of emissivity setting, typical for optical pyrometry. Nevertheless, this type of temperature measurement allows very precise relative temperature control within a single experiment (typically, Fig. 2b).

Localized surface potential variations were captured simultaneously with SEM imaging using the NenoVision LiteScope 1.0. We employed FM-KPFM in a single-pass, tapping mode, tip-stationary regime using NenoProbe conductive tips enhanced with a 200 nm Cu and 20 nm Au coating for improved sensitivity. The topography and potential feedback loops were controlled by a Zurich Instruments HF2LI 50 MHz Lock-in amplifier. The Phase-Locked Loop (PLL) maintained the first resonance frequency at approximately 25 kHz for topographic feedback. For the Kelvin probe signal, the sideband of the second resonance frequency (100 kHz) was utilized, with an AC excitation amplitude of 6 V (peak voltage amplitude) applied to the tip. The bias feedback loop nullified the contact potential difference ($V_{CPD}$) to obtain high-resolution measurements of the surface potential during the surface wave propagation. To avoid artifacts potentially created by the varying tip-sample distance, the temperature-dependence measurements were conducted on the same ROI on the sample while maintaining a fixed tip position. Under these controlled conditions, variations in $V_{CPD}$ are attributed solely to the adsorbate-induced work function differences.

Spatially resolved reaction–diffusion dynamics were simulated using the Metropolis Monte Carlo (MMC) method.[37] To overcome the computational limitations associated with low-energy processes often excluded from the traditional kinetic Monte Carlo method,[37–40] we implemented parallel computing using graphics processing unit (GPU).[41] The simulated area consisted of 2048x2048 surface sites, where each site interacts with four neighboring sites, affecting nearby diffusion and desorption processes. To ensure statistical independence, each GPU core subsequently computes a subarea of 4x4 sites, leaving four sites between two simultaneously processed sites.

The model incorporates CO adsorption, desorption and surface diffusion, dissociative $O_2$ adsorption yielding atomic oxygen, the surface reaction $CO + O \rightarrow CO_2$, subsequent $CO_2$ desorption, and the kinetics of the Pt adsorbate-induced surface phase transition. Event probabilities follow the Arrhenius relation, and local lateral interactions were accounted for by adjusting the binding energies based on the occupancy of the four neighboring sites. Furthermore, to accommodate the rapid propagation of reaction fronts and island nucleation, simulation considers edge adsorption and island-specific enhancement factors.

## ASSOCIATED CONTENT

**Data availability**

The data underlying this study are openly available at XXX.

**Supporting Information**

The Supporting Information is available free of charge at XXX

- Description of supplementary movies, comparison of simulated SE images and experiments under different imaging conditions, additional SEM data analysis.
- Video sequence of operando SEM showing spiral and planar reaction waves on single-crystal Pt(110) during catalytic CO oxidation (Movie M1)
- Video obtained during correlative SEM-KPFM on polycrystalline Pt (Movie M2)
- Adsorbate-induced contrast reversal during an AFM-tip approach to the polycrystalline Pt surface (Movie M3)
- Reaction-diffusion model video sequence showing the spiraling patterns of CO and oxygen waves on platinum (Movie M4)

**Notes**

This manuscript was previously submitted to a preprint server: K.V., M. P., A. O., Z.-J. W., T. Š., P. B., M. K. Work-Function-Resolved Imaging of Relaxation Oscillations and Chemical Spillover in CO Oxidation over Platinum Surfaces. 2026, arXiv:2507.07479v1, 10.48550/arXiv.2507.07479 (accessed July 10, 2025). The authors declare no competing financial interest.

## ACKNOWLEDGMENTS

This research was supported by Quantum materials for applications in sustainable technologies (QM4ST)─project No. CZ.02.01.01/00/22_008/0004572 under OP JAK, call Excellent Research; the Ministry of Education, Youth and Sports of the Czech Republic (LM2023051)